\documentclass[fleqn,10pt]{wlscirep}
\usepackage[utf8]{inputenc}
\usepackage[T1]{fontenc}
\usepackage{colortbl,array,xcolor}
\title{Particle-dependent radioluminescence color response of Eu-doped CaF\texorpdfstring{$_2$}{2} associated with energy-loss density}

\author[1,*]{T. Iida}
\author[2]{M. Yoshino}
\author[1]{T. Moriguchi}
\author[1]{A. Ozawa}
\author[1]{S. Sugi}
\author[1]{M. Mikawa}
\author[1]{M. Mitsui}
\author[1]{H. Kobayashi}
\author[1]{Y. Saito}
\author[1]{R. Iwamoto}
\author[3]{K. J. Kim}
\author[3]{K. Kamada}
\affil[1]{University of Tsukuba, Institute of Pure and Applied Sciences, Tsukuba, 305-8571, Japan}
\affil[2]{Tohoku University, Institute for Materials Research, Sendai, 980-8577, Japan}
\affil[3]{Tohoku University, New Industry Creation Hatchery Center, Sendai, 980-8579, Japan}

\affil[*]{tiida@hep.px.tsukuba.ac.jp}

\begin{abstract}
Europium-doped calcium fluoride (Eu:CaF$_2$) exhibits characteristic blue Eu$^{2+}$ and red Eu$^{3+}$ emissions. In our previous study, the Eu$^{3+}$/Eu$^{2+}$ radioluminescence intensity ratio under $\alpha$-particle irradiation was approximately twice that under X-ray irradiation. Here, we investigate the relationship between this emission ratio and energy-loss density using H, Li, C, and Ni ion beams produced by a tandem accelerator. Radioluminescence spectra were measured for each ion species, and the effective range-averaged energy-loss density was evaluated using SRIM simulations, spanning approximately 18 to 5900~keV/$\mu$m. The Eu$^{3+}$/Eu$^{2+}$ emission-intensity ratio increased systematically from H to C ions by approximately a factor of two, whereas no further systematic increase was observed for Ni ions at substantially higher energy-loss densities. Because the energy-loss density was varied mainly by changing the ion species, its effect cannot be fully separated from that of ion-specific track structure. The X-ray result, included for qualitative comparison, showed a lower emission ratio than the ion-beam measurements, although its effective energy-loss density was estimated only approximately. These results demonstrate a systematic variation in the emission color of Eu:CaF$_2$ across different ion-beam irradiation conditions and provide baseline data for future studies of wavelength-resolved particle discrimination.
\end{abstract}

\begin{document}

\flushbottom
\maketitle
%
%
\thispagestyle{empty}


\section*{Introduction}

Scintillators are functional materials that convert ionizing radiation into visible light and are used extensively in particle and nuclear physics, medical imaging, and environmental radiation monitoring. Europium-doped calcium fluoride (Eu:CaF$_2$) is a representative inorganic scintillator with a light yield of approximately 19,000 photons/MeV, good chemical stability, non-hygroscopicity, and high radiation resistance.\cite{Nikl2000,Nikl2006} Its X-ray-excited luminescence properties, including the absolute luminescence efficiency, have also been quantitatively characterized.\cite{Michail2019Crystals} The relatively low effective atomic number of the CaF$_2$ host ($Z_{\mathrm{eff}}=16.5$) favors Compton scattering relative to photoelectric absorption in the low-energy $\gamma$-ray region. These properties, particularly the relatively low effective atomic number, have recently motivated the development of an omnidirectional Compton camera using Eu:CaF$_2$ to visualize environmental $\gamma$ rays below 250~keV.\cite{KATAGIRI2021}

Eu:CaF$_2$ has also been employed in rare-event searches, including searches for dark matter and neutrinoless double-beta decay, because large crystals can be produced and operated with low radioactive background.\cite{TOVEY1998,Shimizu2005,Umehara2008} Such experiments require effective particle-identification capability to distinguish potential signal events from radioactive and cosmic-ray backgrounds. Many scintillators achieve this separation through pulse-shape discrimination (PSD), which exploits particle-dependent differences in scintillation decay time. Advanced analysis methods, including machine-learning approaches, can further improve PSD performance in scintillators with particle-dependent waveform features.\cite{Yoshino2023} In Eu:CaF$_2$, however, the difference between the decay profiles induced by $\alpha$ particles and by electron recoils from $\beta/\gamma$ radiation is relatively modest, limiting the discrimination performance compared with that of representative scintillators used for PSD. Statistical fluctuations in the detected light can further degrade the discrimination capability, particularly at low energies and under room-temperature operation.\cite{TOVEY1998,Oguri2010}

The linear energy transfer (LET) of incident radiation influences the local density of electron--hole pairs and excitons and their subsequent relaxation and energy-transfer processes. In general, scintillation light yield exhibits a nonlinear response with increasing LET because of excitation-density effects, and this behavior is often described by Birks' law.\cite{Birks1964} When a high-LET charged particle such as a heavy ion traverses a scintillator, the high excitation density near the particle track can cause local quenching and alter the emission dynamics. This is most familiar as the reduced light yield per unit deposited energy observed for $\alpha$-particles relative to $\beta$ or $\gamma$ rays in many inorganic scintillators, commonly expressed as an $\alpha/\beta$ ratio below unity.\cite{Birks1964,TOVEY1998} Whereas Birks' law is commonly used to describe the nonlinear light-yield response, the present study focuses on how excitation density is reflected in the emission spectrum itself, namely in the relative intensities of the two Eu emission channels. In Eu:CaF$_2$, the formation and evolution of self-trapped excitons (STEs), as well as energy transfer to rare-earth activators, are expected to depend on the local excitation density.\cite{CaF2STE1998}

An additional feature of Eu:CaF$_2$ is the coexistence of divalent and trivalent europium ions. Eu$^{2+}$ acts as a blue emission center near 420~nm through an allowed $5d$--$4f$ transition, whereas Eu$^{3+}$ produces characteristic line emission in the 590--700~nm range through intra-$4f$ transitions. Spectroscopic studies have characterized the coexistence of these charge states and shown that their relative populations in synthetic CaF$_2$ are sensitive to dopant concentration and crystal-growth conditions.\cite{Barbin1996,Yu2019OptExpress,Kumar2024Matpr} The local environments of Eu$^{3+}$ ions are also closely associated with charge-compensating defect structures in the CaF$_2$ lattice.\cite{Cirillo-Penn1991} Recently, the Eu$^{3+}$/Eu$^{2+}$ emission-intensity ratio was reported to differ between $\alpha$-particle and X-ray irradiation.\cite{Iida2025SciRep} However, few studies have quantitatively examined this spectral variation in terms of LET, and the relationship between the emission spectrum and energy-loss density remains insufficiently understood. Such a spectral response may provide a particle-identification observable based on emission wavelength, or color, complementary to conventional timing information. Related wavelength-based particle discrimination has also been demonstrated in Eu-doped LiCaI scintillators.\cite{Yoshino2023PTEP}

Despite this potential, practical detector development with Eu:CaF$_2$ has largely emphasized device geometry, light collection, and operational stability, while particle discrimination has continued to rely mainly on conventional waveform analysis. Clarifying the relationship between energy-loss density and the balance between the two Eu emission channels is therefore important for establishing a spectroscopic approach to particle identification.

In this study, we systematically investigated the relationship between energy-loss density and the emission spectrum of Eu:CaF$_2$. Our previous work identified measurements with heavy-ion beams covering a wide range of $dE/dx$ as a necessary next step,\cite{Iida2025SciRep} and the present study addresses that requirement. Specifically, we irradiated the crystal with hydrogen (H), lithium (Li), carbon (C), and nickel (Ni) ion beams using the 6~MV Pelletron tandem accelerator at the University of Tsukuba\cite{Sasa2015Facility,Sasa2016HIAT} and measured the radioluminescence spectra over a broad range of energy-loss densities.

\section*{Experimental setup}

The Eu:CaF$_2$ scintillator crystal used in this study was identical to that employed in our previous work\cite{Iida2025SciRep} (Fig.~1, left). The Eu concentration was 3~mol\%, and the crystal dimensions were $11.5~\mathrm{mm} \times 11.5~\mathrm{mm} \times 6.5~\mathrm{mm}$. All surfaces of the crystal were mirror-polished. As shown in the left panel of Fig.~1, the crystal exhibits reddish-purple luminescence under ultraviolet irradiation.
Figure~1 (right) shows the radioluminescence (RL) spectra of the crystal under $\alpha$-particle (red) and X-ray (blue) irradiation, reproduced from Ref.~\cite{Iida2025SciRep}. The 3~mol\% Eu concentration was selected because the Eu$^{2+}$ and Eu$^{3+}$ emission intensities are comparable, facilitating direct comparison of the two emission components. As shown in the right panel of Fig.~1, the relative intensities of the Eu$^{2+}$ emission near 420~nm and the Eu$^{3+}$ emission at 590--700~nm differ significantly between X-ray and $\alpha$-particle excitation, resulting in a clear difference in the Eu$^{3+}$/Eu$^{2+}$ emission-intensity ratio.

\begin{figure}[ht]
\centering
\includegraphics[width=0.8\linewidth]{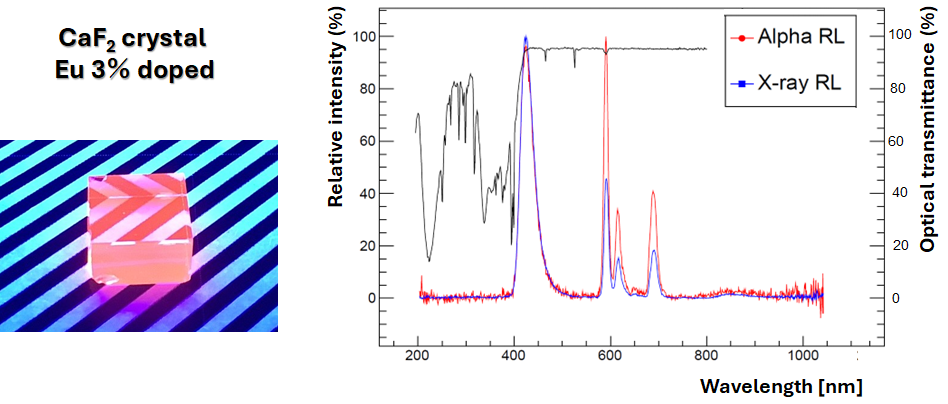}
\caption{(Left) Photograph of the Eu:CaF$_2$ crystal grown, processed, and used in this study under UV illumination (black light). (Right) Radioluminescence spectra of the 3~mol\% Eu-doped CaF$_2$ crystal used in this study under $\alpha$-particle (red) and X-ray (blue) irradiation, reproduced from Ref.~\cite{Iida2025SciRep}. The black curve shows the absolute optical transmittance of the crystal (right axis).}
\label{fig:crystal}
\end{figure}

Ion-beam irradiation experiments were performed using the 6~MV Pelletron tandem accelerator at the University of Tsukuba. The accelerator can provide a wide range of ion species, from hydrogen (H) to gold (Au). In the present study, four ion species were selected: hydrogen (H), lithium (Li), carbon (C), and nickel (Ni). Table~1 summarizes the irradiation conditions and the corresponding beam energies at the crystal surface, as determined by simulation or calculation. To maintain a consistent irradiation area, an aluminum collimator with an 8~mm-diameter aperture was placed in front of the crystal for both the atmospheric and vacuum irradiation setups. The beam current was kept as low as practicable, and the beam intensity was adjusted so that the count rate measured with a plastic scintillator coupled to a photomultiplier tube (PMT) remained below a few $\times 10^{6}$ counts/s.

For H and Li ion irradiations, the beam was extracted into air through a 125-$\mu$m-thick Kapton window installed at the end of the beamline before reaching the crystal, as illustrated in the left panel of Fig.~2. The beam energy at the crystal surface was therefore lower than the initial accelerator energy because of energy loss in the Kapton window. The energy loss and energy straggling in the Kapton window were evaluated using the LISE++ software.\cite{LISEpp} The resulting mean incident energies and their energy spreads (1$\sigma$ standard deviations) are summarized in Table~1. The distance between the Kapton window and the crystal was within a few millimeters, and the corresponding energy loss in air was neglected. Scintillation light was collected from the face opposite to the beam-incident surface using a 400-$\mu$m-core-diameter multimode optical fiber coupled to the crystal with optical grease (Momentive, TSK5353). The light-collection geometry was kept fixed for all measurements. The fiber was connected to a spectrograph (Andor, Shamrock 163), and the dispersed light was recorded as wavelength spectra using a CCD camera (Andor, iDus 420).

For C and Ni ions, which could not be extracted through the Kapton window because of their much larger stopping power, irradiation was performed inside a vacuum chamber installed on the beamline, as illustrated in the right panel of Fig.~2. To vary the incident energies of the C and Ni beams, aluminum foils were placed immediately upstream of the crystal. The energy loss and energy straggling in the aluminum foils were evaluated using LISE++, and the resulting mean incident energies and their energy spreads (1$\sigma$ standard deviations) are summarized in Table~1. The crystal was sandwiched between aluminum plates with an 8-mm-diameter aperture, which also served as the beam collimator, and was mounted on a target ladder, as shown in the photograph in Fig.~2.

\begin{figure}[ht]
\centering
\includegraphics[width=1.0\linewidth]{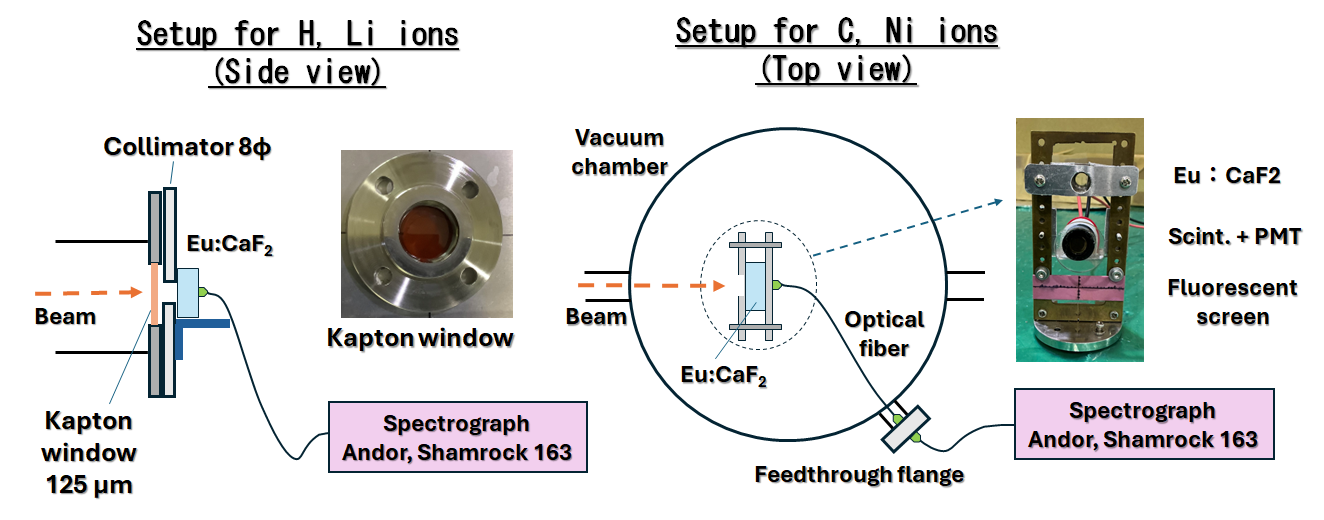}
\caption{Experimental setup for radioluminescence measurements under ion-beam irradiation. (Left) Side-view schematic of the atmospheric irradiation setup for H and Li ions, together with a photograph of the Kapton window. (Right) Top-view schematic of the vacuum irradiation setup for C and Ni ions, together with a photograph of the target ladder holding the Eu:CaF$_2$ crystal between aluminum plates with an 8-mm-diameter aperture.}
\label{fig:RL}
\end{figure}

In addition to the crystal, a beam-monitoring detector (a plastic scintillator coupled with a PMT) and an alumina fluorescent screen (Desmarquest, AF995R) were mounted at different positions on the ladder~\cite{Moriguchi2024UTTAC}.
The ladder position could be adjusted remotely, allowing these components to serve as beam intensity and position monitors as needed. 
The scintillation light from the crystal was carried by the same type of $400\ \mu\text{m}$ optical fiber (optically coupled with TSK5353 grease) to an FC-based vacuum feedthrough flange equipped with FC adapters on both sides. FC-connectorized fibers were connected to these adapters to transmit the light out of the chamber and route it to the same spectroscopic system (spectrograph and CCD camera) used in the atmospheric measurements.

\section*{Results}

\begin{figure}[ht]
\centering
\includegraphics[width=0.7\linewidth]{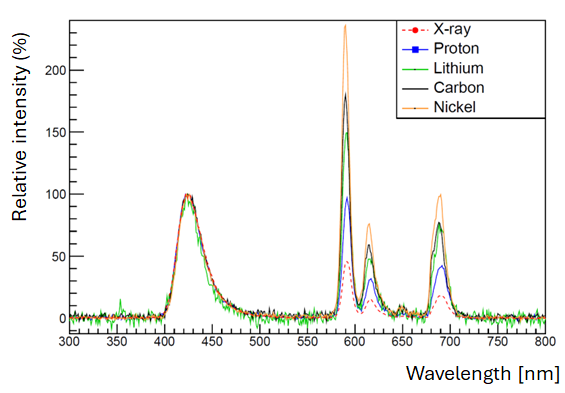}
\caption{Radioluminescence spectra of the Eu:CaF$_2$ crystal obtained in the present tandem-accelerator experiment under H (blue), Li (green), C (black), and Ni (orange) ion irradiation. The spectrum under X-ray irradiation reported in our previous study is shown in red for comparison.\cite{Iida2025SciRep} Each spectrum was normalized such that the intensity of the Eu$^{2+}$ emission peak near 420~nm was set to 100.}
\label{fig:RL1}
\end{figure}

Figure~\ref{fig:RL1} shows the radioluminescence spectra of the Eu:CaF$_2$ crystal under irradiation with various ion beams generated by the tandem accelerator. The emission peak near 420~nm is attributed to the $5d$--$4f$ transition of Eu$^{2+}$, whereas the emission bands in the 590--700~nm range originate from the intra-$4f$ transitions of Eu$^{3+}$. The horizontal axis represents the emission wavelength, and the vertical axis represents the relative emission intensity. All spectra obtained under ion-beam irradiation were normalized such that the Eu$^{2+}$ emission peak intensity was set to 100. The spectra are color-coded according to the incident ion species: H (blue), Li (green), C (black), and Ni (orange). For comparison, the radioluminescence spectrum under X-ray irradiation reported in our previous study~\cite{Iida2025SciRep} is also shown in red.

As reported in our previous study, the Eu$^{3+}$/Eu$^{2+}$ emission-intensity ratio differs between X-ray and $\alpha$-particle irradiation.\cite{Iida2025SciRep} Because the spectra are normalized to the Eu$^{2+}$ emission peak, changes in the relative Eu$^{3+}$ emission intensity directly reflect changes in the Eu$^{3+}$/Eu$^{2+}$ ratio. As shown in Fig.~\ref{fig:RL1}, the relative Eu$^{3+}$ emission intensity increases from H to Li and C irradiation, while no clear further increase is observed for Ni ions. The Eu$^{3+}$/Eu$^{2+}$ emission-intensity ratio was defined as the ratio of the maximum Eu$^{3+}$ peak intensity near 590~nm to the Eu$^{2+}$ peak intensity near 420~nm. Peak intensities rather than integrated emission areas were used because small run-to-run variations in the spectral baseline could introduce additional systematic uncertainty when integrating over broad wavelength ranges. The peak-intensity ratio is less sensitive to such baseline variations and was therefore adopted as the primary observable in this study. The resulting values, normalized to that for H-ion irradiation, are summarized in Table~\ref{tab:summary}. The relationship between this ratio and the energy-loss density of each radiation type is examined below.

\begin{figure}[ht]
\centering
\includegraphics[width=\linewidth]{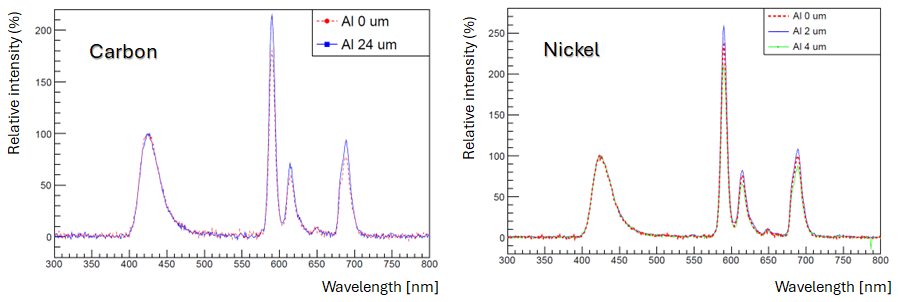}
\caption{Radioluminescence spectra of the Eu:CaF$_2$ crystal measured at different incident ion energies. The ion energies were varied by inserting aluminum (Al) foils into the beam path. (Left) C-ion spectra recorded without an Al foil (red) and with a 24-$\mu$m-thick Al foil (blue). (Right) Ni-ion spectra recorded without an Al foil (red) and with 2-$\mu$m-thick (blue) or 4-$\mu$m-thick (green) Al foils.}
\label{fig:Alfoil}
\end{figure}

Because the energy-loss density depends not only on the ion species but also on the incident ion energy, additional measurements were performed at different beam energies for the same ion species by inserting aluminum foils of different thicknesses into the beam path, as described in the previous section. Figure~\ref{fig:Alfoil} shows the radioluminescence spectra obtained under these conditions. The left panel shows the spectra for C-ion irradiation without an aluminum foil (red) and with a 24-$\mu$m-thick aluminum foil (blue). The right panel shows the spectra for Ni-ion irradiation without an aluminum foil (red) and with 2-$\mu$m- (blue) and 4-$\mu$m-thick (green) aluminum foils.
Modest changes in the Eu$^{3+}$ emission intensity relative to the Eu$^{2+}$ emission were observed as the incident ion energy was varied. These measurements provide a comparison of the spectral response at different energy-loss densities within the same ion species.

\begin{figure}[ht]
\centering
\includegraphics[width=0.9\linewidth]{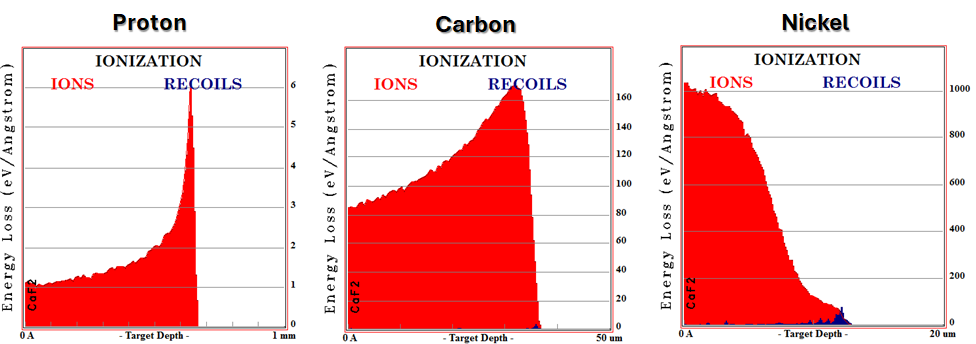}
\caption{Representative SRIM simulations of the depth-dependent energy loss of H, C, and Ni ions in CaF$_2$. The red and blue regions represent energy loss due to electronic ionization and nuclear recoils, respectively. The incident energies are 11.3 MeV for H, 42 MeV for C, and 72 MeV for Ni. Note that the horizontal and vertical scales differ among the three panels. Under the present irradiation conditions, electronic ionization is the dominant energy-loss process.}
\label{fig:SRIM}
\end{figure}

To quantify the energy-loss density for each irradiation condition, the stopping of ions in CaF$_2$ was simulated using the SRIM code~\cite{SRIM}. For each ion species and incident energy, 1000 ion trajectories were simulated to obtain the mean projected range and the depth-dependent energy-loss profile. Representative results for H, C, and Ni ions are shown in Fig.~\ref{fig:SRIM}. Under the present irradiation conditions, electronic energy loss through ionization dominates over the contribution from nuclear recoils. For comparison among the different irradiation conditions, an effective range-averaged energy-loss density, $\langle dE/dx\rangle_{\mathrm{eff}}$, was calculated as
\[
\langle dE/dx\rangle_{\mathrm{eff}} = \frac{E_0}{R_p},
\]
where $E_0$ is the incident ion energy at the crystal surface and $R_p$ is the mean projected range obtained from SRIM. This quantity was used as an LET-related descriptor rather than as the local stopping power along the ion track. The energy spreads arising from straggling in the Kapton window and aluminum foils were propagated to the calculated $\langle dE/dx\rangle_{\mathrm{eff}}$ values, as summarized in Table~\ref{tab:summary}. The range straggling obtained from the SRIM simulations was small compared with the mean projected range for the ion species and energies investigated here.

For X-ray irradiation, an analogous effective energy-loss density was estimated from the range of the secondary electrons rather than from the X-ray attenuation length. Since the X-ray tube was operated at 40~kV, a representative photon energy of approximately 20~keV was assumed. For Ca K-shell photoelectric absorption, whose binding energy is approximately 4.04~keV,\cite{Deslattes2003} this corresponds to a photoelectron kinetic energy of approximately 16~keV. The continuous-slowing-down approximation (CSDA) range of a 16-keV electron in CaF$_2$, estimated by interpolation of the NIST ESTAR electron-range data,\cite{NISTESTAR} is approximately 2.5~$\mu$m. The corresponding effective range-averaged energy-loss density was therefore estimated as
\[
\left\langle dE/dx \right\rangle_{\mathrm{eff}}
\simeq \frac{16~\mathrm{keV}}{2.5~\mu\mathrm{m}}
\simeq 6.4~\mathrm{keV/\mu m}.
\]
This value was used as an approximate LET-related descriptor for X-ray-induced secondary electrons. Because the X-ray source produces a broad photon-energy distribution and multiple secondary-electron processes contribute to energy deposition, this value should be regarded as a representative quantity rather than a uniquely defined stopping power.
The resulting $\langle dE/dx\rangle_{\mathrm{eff}}$ values for all irradiation conditions are summarized in Table~\ref{tab:summary}.

\begin{table}[ht]
\centering
\begin{tabular}{c|c|c|c|c|c} \hline
\rowcolor[gray]{0.8}%
Beam & Al foil & Incident energy & Range & $\langle dE/dx\rangle_{\mathrm{eff}}$ & Relative Eu$^{3+}$/Eu$^{2+}$ \\
\hline
X-ray & - & $\sim$16 keV (photoelectron) & 2.5 $\mu$m  & 6.4 keV/$\mu$m & 0.47 $\pm$0.05 \\
$\alpha$-particle & - & 5.5 MeV & 20.2 $\mu$m  & 272.3 keV/$\mu$m & 1.11 $\pm$0.11\\
H & - & 11.25 $\pm$0.03 MeV & 640 $\mu$m & 17.66 $\pm$0.05 keV/$\mu$m & 1.00 $\pm$0.10\\
Li & - & 1.08 $\pm$0.25 MeV & 2.45 $\mu$m & 441 $\pm$102 keV/$\mu$m & 1.58 $\pm$0.16\\
C & 0 $\mu$m & 42 MeV & 36.0 $\mu$m & 1167 keV/$\mu$m & 1.90 $\pm$0.19\\
C & 24 $\mu$m & 23.5 $\pm$0.17 MeV & 17.1 $\mu$m  & 1374 $\pm$10 keV/$\mu$m & 2.26 $\pm$0.23\\
Ni & 0 $\mu$m & 72 MeV & 12.3 $\mu$m & 5854 keV/$\mu$m & 2.47 $\pm$0.25\\
Ni & 2 $\mu$m & 55.23 $\pm$0.09 MeV & 10.7 $\mu$m  & 5168 $\pm$8 keV/$\mu$m & 2.69 $\pm$0.27\\
Ni & 4 $\mu$m & 39.30 $\pm$0.12 MeV & 9.08 $\mu$m & 4328 $\pm$13 keV/$\mu$m & 2.21 $\pm$0.22\\
\hline
\end{tabular}
\caption{\label{tab:summary} Summary of irradiation conditions and results: radiation type, Al foil thickness, incident energy, mean projected range, effective range-averaged energy-loss density $\langle dE/dx\rangle_{\mathrm{eff}}$, and Eu$^{3+}$/Eu$^{2+}$ emission-intensity ratio normalized to H-ion irradiation. Ranges were calculated using SRIM for ions and ESTAR for a representative X-ray-induced photoelectron. Incident-energy uncertainties indicate 1$\sigma$ spreads from straggling in the Kapton window or Al foils, calculated using LISE++. The intrinsic accelerator energy spread ($\Delta E/E\sim10^{-4}$) is negligible compared with these spreads; accelerator energy-calibration uncertainties are not included. Uncertainties in $\langle dE/dx\rangle_{\mathrm{eff}}$ were estimated with the SRIM projected ranges held fixed. Emission-ratio uncertainties represent 10\% relative measurement reproducibility.}
\end{table}

\begin{figure}[ht]
\centering
\includegraphics[width=0.8\linewidth]{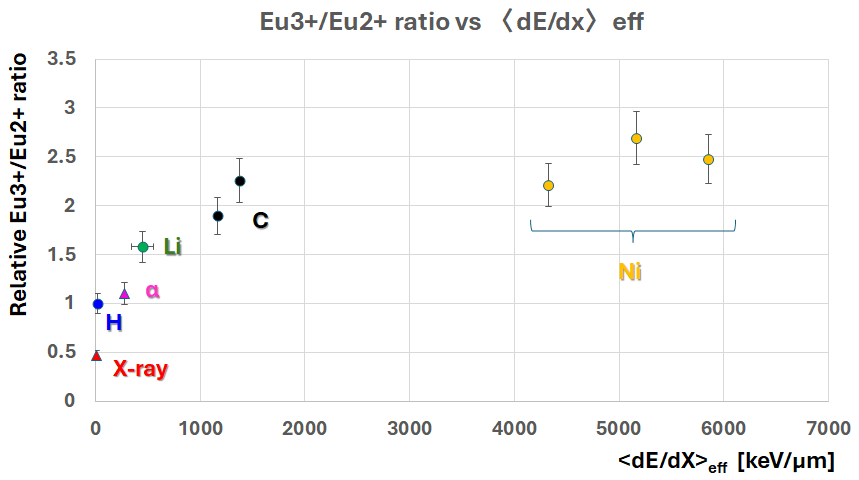}
\caption{Relationship between the effective range-averaged energy-loss density, $\langle dE/dx\rangle_{\mathrm{eff}}$, and the relative Eu$^{3+}$/Eu$^{2+}$ emission-intensity ratio, normalized to that for H-ion irradiation. Circles represent the ion-beam measurements obtained in the present study, while triangles indicate the X-ray and $\alpha$-particle results from our previous study~\cite{Iida2025SciRep}. The marker colors distinguish the radiation species. The X-ray point is shown for qualitative reference only. Horizontal error bars represent the energy spreads (1$\sigma$) due to straggling in the Kapton window or aluminum foils, converted to $\langle dE/dx\rangle_{\mathrm{eff}}$ with the mean projected ranges fixed at their nominal SRIM values. Vertical error bars represent a 10\% relative uncertainty conservatively estimated from the reproducibility of repeated H-ion measurements.}
\label{fig:LET}
\end{figure}

Figure~\ref{fig:LET} shows the relationship between the effective range-averaged energy-loss density, $\langle dE/dx\rangle_{\mathrm{eff}}$, and the relative Eu$^{3+}$/Eu$^{2+}$ emission-intensity ratio, normalized to that for H-ion irradiation. The ion-beam data obtained in the present study are shown as circles, while the X-ray and $\alpha$-particle data reported in our previous study~\cite{Iida2025SciRep} are shown as triangles for reference. The marker colors distinguish the radiation species. The X-ray point is included only for qualitative comparison because LET is not uniquely defined for photon irradiation; its plotted value represents an effective energy-loss density estimated for a representative secondary photoelectron. Multiple data points are shown for C and Ni ions because measurements were performed at different incident energies using aluminum foils of different thicknesses. The horizontal error bars represent the energy spreads due to straggling in the Kapton window or aluminum foils, converted to uncertainties in $\langle dE/dx\rangle_{\mathrm{eff}}$ using fixed mean projected ranges, as described in Table~\ref{tab:summary}.

The statistical uncertainty of the Eu$^{3+}$/Eu$^{2+}$ emission-intensity ratio was negligible owing to the large number of detected photons. To assess the measurement reproducibility, H-ion irradiation was repeated three times under nominally identical conditions, including remounting of the crystal and small variations in the beam intensity. The relative variation among the three measurements, defined as $(R_{\max}-R_{\min})/\bar{R}$, where $R$ denotes the measured Eu$^{3+}$/Eu$^{2+}$ ratio, was approximately 10\%. A relative uncertainty of 10\% was therefore assigned to all emission ratios as a conservative estimate of the experimental uncertainty associated with measurement reproducibility, as reported in Table~\ref{tab:summary} and shown by the vertical error bars in Fig.~\ref{fig:LET}.

As shown in Fig.~\ref{fig:LET}, the relative Eu$^{3+}$/Eu$^{2+}$ emission ratio increases systematically from H (1.00) through Li (1.58) to C ions (1.90--2.26) with increasing $\langle dE/dx\rangle_{\mathrm{eff}}$. A similar tendency is observed within the C-ion measurements: reducing the incident energy using a 24-$\mu$m-thick Al foil increased $\langle dE/dx\rangle_{\mathrm{eff}}$ from 1167 to 1374~keV/$\mu$m and the emission ratio from 1.90 to 2.26, although this difference is comparable to the estimated experimental uncertainty. For Ni ions, which have substantially larger $\langle dE/dx\rangle_{\mathrm{eff}}$ values of approximately 4300--5900~keV/$\mu$m, the emission ratios (2.21--2.69) show no further systematic increase, with variations comparable to the estimated experimental uncertainty. The $\alpha$-particle result from our previous study (1.11) lies between the H and Li points, closer to the H value, whereas the X-ray point (0.47) lies below the ion-beam trend; both are included for reference, with the X-ray point used only for qualitative comparison as described above. Overall, the Eu$^{3+}$/Eu$^{2+}$ emission ratio increases across the ion-beam conditions up to approximately 1400~keV/$\mu$m in $\langle dE/dx\rangle_{\mathrm{eff}}$, while no clear further increase is observed at higher values.

\section*{Discussion}

Whereas our previous study compared only X-ray and $\alpha$-particle irradiation~\cite{Iida2025SciRep}, the present work covers four ion species (H, Li, C, and Ni) spanning about two and a half orders of magnitude in $\langle dE/dx\rangle_{\mathrm{eff}}$, from approximately 18 to 5900~keV/$\mu$m (Table~\ref{tab:summary}). The Eu$^{3+}$/Eu$^{2+}$ ratio increases systematically from H to C ions, whereas no further systematic increase is observed for Ni ions. Two limitations should be kept in mind when interpreting this trend. First, $\langle dE/dx\rangle_{\mathrm{eff}}$ is a range-averaged quantity and does not resolve the strong variation of the local energy-loss density along an ion track; a local or dose-averaged LET, together with a detailed treatment of track structure, would provide a more rigorous description. The Li ions, in particular, entered the crystal at only 1.1 MeV after substantial energy loss in the Kapton window, resulting in a relatively large energy spread and uncertainty in $\langle dE/dx\rangle_{\mathrm{eff}}$. Second, because the broad range of $\langle dE/dx\rangle_{\mathrm{eff}}$ was obtained primarily by changing the ion species, the effects of energy-loss density and of ion-specific track structure cannot be completely separated. The present results therefore indicate that the emission spectrum of Eu:CaF$_2$ is sensitive to the energy-loss conditions of the incident particle, rather than demonstrating a unique dependence on a single LET parameter.
We therefore discuss possible experimental effects, excitation-density-dependent emission mechanisms, and other processes that may contribute to the observed spectral response.

Before discussing possible mechanisms, we consider two experimental effects that could mimic the observed trend. The first is self-absorption of the blue Eu$^{2+}$ emission. Because the scintillation light is collected from the face opposite to the irradiated surface and the excitation depth ranges from approximately 2.5~$\mu$m for Li to 640~$\mu$m for H (Table~\ref{tab:summary}), the optical path length to the collection face differs by at most approximately 0.64~mm among the ion species. In our previous study, two samples with thicknesses of 1 and 5~mm, cut from a single 3~mol\% Eu:CaF$_2$ crystal grown independently of the present one, gave Eu$^{3+}$/Eu$^{2+}$ ratios under $\alpha$-particle irradiation that differed by less than 3\% despite the 4~mm difference in optical path length (Fig.~5, right, of Ref.~\cite{Iida2025SciRep}). Assuming exponential attenuation, this corresponds to an effect of less than 1\% for a 0.64~mm path difference, far smaller than both the variation observed here and the measurement reproducibility (10\%). The second is the difference between the optical configurations used for the atmospheric (H and Li) and vacuum (C and Ni) measurements; in the latter, the light additionally passed through a 1.0-m-long silica fiber and a vacuum feedthrough with two additional connections. The specified fiber attenuation corresponds to a transmission loss of the order of 1\% over 1~m, its wavelength dependence between 420 and 590--700~nm is expected to be smaller still, and wavelength-independent connection losses cancel in the intensity ratio. More importantly, the ratio already increases by approximately 58\% from H to Li ions, both of which were measured in the same atmospheric configuration. These effects are therefore unlikely to account for the observed trend, although a small wavelength-dependent difference between the two configurations could shift the absolute scale of the C- and Ni-ion data relative to the H- and Li-ion data.

One possible interpretation is a competition between the two Eu-related emission channels for the excitation energy generated in the host. In Eu:CaF$_2$, Eu is incorporated predominantly as Eu$^{2+}$, which isovalently substitutes for Ca$^{2+}$, whereas Eu$^{3+}$ forms a minority population associated with charge-compensating defects, whose relative fraction increases with dopant concentration~\cite{Yu2019OptExpress,Hamers1982,Cirillo-Penn1991,Kumar2024Matpr}. Electron--hole pairs and self-trapped excitons generated in the host can relax through several pathways involving Eu-related and defect-related centres~\cite{CaF2STE1998}. At low excitation density, as for X-rays and low-LET ions, the abundant Eu$^{2+}$ centres may accommodate a larger fraction of the excitation energy, resulting in a relatively small Eu$^{3+}$/Eu$^{2+}$ ratio. Along denser ion tracks, the relative contribution of the Eu$^{3+}$ emission may increase through two non-exclusive routes: saturation or other limitations of the Eu$^{2+}$-related pathway, and non-radiative quenching at high excitation density, analogous to that described by Birks' law~\cite{Birks1964}, if it suppresses the Eu$^{2+}$-related emission more strongly than the Eu$^{3+}$-related emission. Either route would increase the ratio even without an absolute enhancement of the Eu$^{3+}$ emission. The absence of a further increase for Ni ions suggests that this change in the relative contributions does not continue indefinitely at the highest excitation densities; possible reasons include saturation of the Eu$^{3+}$-related pathway itself or quenching that affects both pathways to a similar extent. Because all spectra were normalized to the Eu$^{2+}$ peak, however, the present data show a change in the relative emission intensities but cannot distinguish an enhancement of the Eu$^{3+}$ emission from a suppression of the Eu$^{2+}$ emission, nor identify the underlying microscopic pathway. The mechanisms above should therefore be regarded as qualitative possibilities.

The Eu-concentration dependence observed in our previous study provides an additional constraint. In a simple model based on depletion or saturation of available Eu$^{2+}$ centres, the spectral response might be expected to weaken with increasing Eu$^{2+}$ concentration. However, the enhancement of the Eu$^{3+}$/Eu$^{2+}$ ratio under $\alpha$-particle irradiation relative to X-ray irradiation was nearly independent of Eu concentration above 1~mol\%~\cite{Iida2025SciRep}. This suggests that Eu$^{2+}$ depletion alone is insufficient to explain the spectral response, at least at the energy-loss density of $\alpha$ particles, and that the quenching discussed above may also contribute. Because the Eu concentration also affects the relative populations and local environments of the Eu$^{2+}$ and Eu$^{3+}$ centres, this argument is not conclusive; measurements of the concentration dependence at higher energy-loss densities would provide a more direct test.

The X-ray point lies below the trend of the ion-beam data. This may partly reflect the different microscopic distribution of energy deposition: X-rays deposit their energy through photoelectrons and subsequent secondary electrons distributed over the irradiated volume, rather than through a single dense ion track. Because the X-ray value of $\langle dE/dx\rangle_{\mathrm{eff}}$ is only a representative quantity for a broad photon spectrum and the X-ray spectrum was obtained in a separate experiment~\cite{Iida2025SciRep}, this comparison remains qualitative. In addition, radiation-induced changes in the Eu charge state cannot be excluded. Soft X-ray irradiation has been shown to induce stable Eu$^{3+}$-to-Eu$^{2+}$ conversion in Eu:CaF$_2$ nanomaterials through electron trapping at Eu$^{3+}$ centres~\cite{Shaidulin2026}, which would lower the Eu$^{3+}$/Eu$^{2+}$ ratio. Whether comparable conversion occurs in the present bulk crystal, under either X-ray or ion irradiation, remains to be examined.

Distinguishing among these possibilities will require further measurements. Absolute, energy-normalized intensities of the Eu$^{2+}$ and Eu$^{3+}$ bands would separate an enhancement of the Eu$^{3+}$ emission from a suppression of the Eu$^{2+}$ emission. Measurements with different ion species at overlapping energy-loss densities, together with wider energy scans within a single species, would help separate the effect of energy-loss density from that of ion-specific track structure, and collecting the emission from a different crystal face would help assess the influence of excitation depth and optical path length. Time-resolved measurements of the two components would be particularly informative because saturation, non-radiative quenching, and energy-transfer processes may produce different changes in the emission amplitudes and decay kinetics. Correlated temporal changes in the Eu$^{2+}$ and Eu$^{3+}$ emissions could provide evidence for coupling between the two emission pathways, although such signatures may also be influenced by carrier trapping, defect-related processes, or changes in the Eu charge state.

\section*{Conclusion and Future Outlook}

The radioluminescence spectra of Eu:CaF$_2$ were investigated under irradiation with H, Li, C, and Ni ion beams generated by a tandem accelerator, and the Eu$^{3+}$/Eu$^{2+}$ emission-intensity ratio was compared with the effective range-averaged energy-loss density, $\langle dE/dx\rangle_{\mathrm{eff}}$, evaluated using SRIM simulations. The ratio increased systematically with increasing $\langle dE/dx\rangle_{\mathrm{eff}}$ from H to C ions, whereas no further systematic increase was observed for Ni ions. Self-absorption and differences between the optical configurations are unlikely to account for the observed trend. The X-ray result, included for qualitative comparison, showed a lower emission ratio than the ion-beam data, although its effective energy-loss density was estimated only approximately from a representative secondary photoelectron.

These results show that the radioluminescence spectrum of Eu:CaF$_2$ varies systematically with the energy-loss conditions of the incident charged particle over a broad range of $\langle dE/dx\rangle_{\mathrm{eff}}$. Because this range was obtained mainly by changing the ion species, the contributions of energy-loss density and ion-specific track structure cannot yet be fully separated, and the present data do not distinguish an enhancement of the Eu$^{3+}$ emission from a suppression of the Eu$^{2+}$ emission. The present study nevertheless provides experimental data for exploring the Eu$^{3+}$/Eu$^{2+}$ emission-intensity ratio as a potential optical indicator of energy-loss conditions in Eu:CaF$_2$ scintillators.

The observed spectral response may also provide baseline information for future studies of wavelength-resolved particle discrimination. For example, a dual-channel optical readout system employing a dichroic mirror or bandpass filters could separately detect the blue Eu$^{2+}$ and red Eu$^{3+}$ emission components using photodetectors such as SiPMs or PMTs. However, the Eu$^{3+}$ intra-$4f$ emission is relatively slow, typically on the millisecond timescale, which may limit the achievable count rate and timing performance of such a system. Future event-by-event measurements with wavelength-resolved detection will be required to evaluate the feasibility and practical particle-discrimination performance of this approach.

\section*{Data Availability}

The datasets generated and/or analyzed during the current study are available from the corresponding author upon reasonable request.

\bibliography{sample}



\section*{Acknowledgements}

The authors thank the University of Tsukuba Tandem Accelerator Complex (UTTAC) for providing access to its accelerator facilities through the shared-use program.
The authors gratefully acknowledge the technical support provided by the staff members of UTTAC.

\section*{Funding}

This study was financially supported by JSPS KAKENHI Grant-in-Aid for Challenging Research (Exploratory) (grant numbers JP25K22021 and JP23K17685), the JSPS Bilateral Program (grant number JPJSBP120252502), the Asahi Glass Foundation, and the Konica Minolta Science and Technology Foundation. The funders played no part in designing the study, conducting the experiment, analysing the results or writing the manuscript.

\section*{Author contributions statement}

Conception: T. Iida.  
Methodology and experiment: T. Iida, M. Yoshino, T. Moriguchi, A. Ozawa.  
Experimental support and accelerator/beamline operation: S. Sugi, M. Mikawa, M. Mitsui, H. Kobayashi, Y. Saito, R. Iwamoto.  
Resources: K. Kamada, K. J. Kim.  
Analysis: T. Iida.  
Writing and editing the manuscript: T. Iida, M. Yoshino.  

All authors reviewed the manuscript and approved the final version.

\section*{Additional information}


\textbf{Competing interests} 

The authors declare no competing interests.



\end{document}